\documentclass[showpacs,twocolumn,superscriptaddress,prb]{revtex4-1}

\usepackage{graphicx}
\usepackage{color}
\begin{document}

\title{Possible nodal superconducting gap in Fe$_{1+y}$(Te$_{1-x}$Se$_{x}$) single crystals from ultra-low temperature penetration depth measurements}

\author{Andrei Diaconu}
\affiliation{Advanced Materials Research Institute - AMRI, University of New Orleans, New Orleans, LA 70148, USA}
\affiliation{Department of Physics, University of New Orleans, New Orleans, LA 70148, USA}

\author{Catalin Martin}
\affiliation{Department of Physics, University of Florida, Gainesville, FL 32611, USA}

\author{Jin Hu}
\affiliation{Department of Physics and Engineering Physics, Tulane University, New Orleans, LA 70118, USA}

\author{Tijiang Liu}
\affiliation{Department of Physics and Engineering Physics, Tulane University, New Orleans, LA 70118, USA}

\author{Bin Qian}
\affiliation{Department of Physics and Engineering Physics, Tulane University, New Orleans, LA 70118, USA}

\author{Zhiqiang Mao}
\affiliation{Department of Physics and Engineering Physics, Tulane University, New Orleans, LA 70118, USA}

\author{Leonard Spinu}
\affiliation{Advanced Materials Research Institute - AMRI, University of New Orleans, New Orleans, LA 70148, USA}
\affiliation{Department of Physics, University of New Orleans, New Orleans, LA 70148, USA}
\email[Electronic address]{LSpinu@uno.edu}

\begin{abstract}
Using a radio frequency tunnel diode oscillator technique, we measured the temperature dependence of the in-plane London penetration depth $\Delta\lambda_{ab}(T)$ in Fe$_{1+y}$(Te$_{1-x}$Se$_{x})$ single crystals, down to temperatures as low as 50 mK. A significant number of samples, with nominal Se concentration $x$=0.36, 0.40, 0.43 and 0.45 respectively, were studied and in many cases we found that $\Delta\lambda_{ab}(T)$ shows an upturn below 0.7 K, indicative of a paramagnetic type contribution. After subtracting the magnetic background, the low temperature behavior of penetration depth is best described by a power law with exponent $n\approx2$ and with no systematic dependence on the Se concentration. Most importantly, in the limit of T$\rightarrow$0, in some samples we observed a narrow region of linear temperature dependence of penetration depth, suggestive of nodes in the superconducting gap of Fe$_{1+y}$(Te$_{1-x}$Se$_{x})$.
\end{abstract}


\maketitle

\section{Introduction}

The iron chalcogenides represent a special class of Fe-based superconductors, with perhaps the simplest layered structure, the so called (11). Superconductivity with critical temperature $T_{c}= 8$K was first reported in the PbO-type structure $\beta$-FeSe~\cite{Hsu08}, and soon thereafter, $T_{c}$ was increased to about 37 K under applied pressure~\cite{Medvedev09}. Initially, this was directly linked to Se deficiencies~\cite{Hsu08}, but later studies~\cite{Hu12} also revealed the sensitivity of the critical temperature to the Fe non-stoichiometry.

The isostructural chalcogenide Fe$_{1+y}$Te is an antiferromagnet, with ($\pi$,0) magnetic wave-vector; upon Te substitution with Se ~\cite{Fang08,Yeh08,Mizuguchi09} it becomes superconductive with an optimum doping level of 50$\%$ Se. Combining several experimental measurements, such as resistivity, Hall effect, magnetic susceptibility, specific heat and neutron scattering, Liu et al.\cite{Liu10} determined the phase diagram of Fe$_{1.02}$(Te$_{1-x}$Se$_{x})$ for Se concentration ranging from un-doped to optimally doped. Although zero transport resistance was observed for all Se concentrations, both specific heat and susceptibility measurements revealed that the bulk superconductivity does not occur until $x\geq$ 0.3 and the maximum T$_{c}\approx$~14 K is obtained for $x\approx$ 0.50~\cite{Liu10,arxiv}. It was also found that with Se doping, the ($\pi$,0) magnetic correlations are suppressed and the ($\pi$,$\pi$) magnetic resonance was observed in the superconducting state for the samples that show bulk superconductivity.

Therefore, because iron pnictides also show superconductivity close to $(\pi,\pi)$ magnetic instabilities, the pairing mechanism in Fe$_{1+y}$(Te$_{1-x}$Se$_{x})$ may very likely be the same as in the FeAs-based compounds. However, the symmetry and the structure of the superconducting gap(s), which are intimately related to the pairing mechanism, are still debated both in the FeAs and, perhaps even more so, in the Fe chalcogenide materials. Two independent reports of scanning tunneling microscopy (STM) seem to suggest a transition from a nodal superconducting gap, in FeSe to a nodeless $s_{\pm}$ gap symmetry in Fe$_{1+y}$(Te$_{1-x}$Se$_{x})$~\cite{Hanaguri10, Song11}. However, specific heat studies reveal isotropic gap behaviour under zero magnetic field~\cite{Hu11} but anisotropic/nodal gaps under magnetic field for optimally doped  Fe(Se, Te) samples~\cite{Zeng10}.
\newline \indent One of the most involved probes for studying Fe$_{1+y}$(Te$_{1-x}$Se$_{x})$ superconductors is the London penetration depth. Measurements of $\lambda(T)$ are directly related to the density of states and provide a powerful tool for investigating low lying quasiparticles energy and, for this very reason, can give valuable hints on superconducting gap function symmetry.
Muon-spin rotation spectrometry ($\mu$-SR)~\cite{Biswas10,Bendele10} and microwave cavity studies~\cite{Takahashi11} showed that superfluid density for $x$=0.50 and $x$=0.41 respectively, is consistent with two gaps with s$\pm$ symmetry. The microwave measurements also found that at low temperature, $\Delta\lambda(T)$ has a nearly quadratic behavior. Similar power law temperature dependence, with exponent $n\approx$ 2, was also reported from radio-frequency tunnel diode oscillator (TDO) data by several groups~\cite{Kim10, Serafin10, Klein10, Cho11}. Most previous TDO studies however, focus on one particular concentration, specially close to the optimal doping, and there seem to be relatively large variations in the magnitude of $\Delta\lambda(T)$ between different measurements. Moreover, we are aware of only one TDO study at temperatures below 0.5 K, performed on Fe$_{1.0}$Te$_{0.44(4)}$Se$_{0.56(4)}$ samples, where the in-plane penetration depth revealed an upturn at low temperatures, attributed to paramagnetic impurities~\cite{Serafin10}.
\newline \indent In this work we present a systematic study of the temperature dependence of the in-plane penetration depth ($\Delta\lambda_{ab}(T)$) in Fe$_{1+y}$(Te$_{1-x}$Se$_{x})$. We measured a significant number of single crystals with different Se concentrations $(x=0.36$, $0.40$, $0.43$ and $0.45)$ and our measurements were extended down to 50 mK in order to better understand the pairing symmetry of this system and its evolution with doping.

\section{Experimental}

Single crystals of Fe$_{1+y}$(Te$_{1-x}$Se$_{x}$), synthesized using the flux technique, with nominal compositions $y=0$ and $x=0.36$, $0.40$, $0.43$ and $0.45$ respectively, were selected from the same batches as those used in Ref.\,\onlinecite{Liu10} for determining the phase diagram. The actual composition of the samples has been shown to slightly differ from the nominal one; an excess of iron up to 2$\%$ (i.e. $y\approx$ 0.02) is observed in most samples. Using magnetic susceptibility and heat capacity measurements a large number of samples with highest superconducting volume fraction were selected for this study. However, in this article we only show data on two samples for each Se concentration. All samples under test are in the shape of rectangular slabs with approximate dimensions of 2$\times$2$\times$0.1 mm$^{3}$.

The temperature dependence of the in-plane penetration depth $\Delta\lambda(T)$ was measured using a tunnel diode oscillator (TDO) technique \cite{Vandegrift75}, incorporated in a dilution refrigerator. A magnetically active sample placed in the ac field generated by the LC tank coil will modify its inductance and consequently the resonant frequency of the TDO circuit. A change in the susceptibility $\Delta\chi$ of the sample will generate a directly proportional change in inductance $\Delta L$ hence, for $\Delta L\ll L$, a proportional shift in resonant frequency\cite{Vannette08} $\Delta f\propto\Delta\chi$.

The susceptibility $\chi$ of a rectangular slab shaped superconductor in Meissner state, under a uniform perpendicular applied magnetic field, was shown to have the following dependence on penetration depth~\cite{Prozorov00, Prozorov06}:

\begin{equation}
\label{eq:eq1}
-4\pi\chi = \frac{1}{1-N}\left[1-\frac{\lambda}{R}\tanh\left(\frac{R}{\lambda}\right)\right]
\end{equation}
where $R$ is an effective dimension of the sample and $N$ is an effective demagnetizing factor both depending on the sample geometry. It follows that at low temperatures $\lambda\ll R$ and therefore the changes in susceptibility $\Delta\chi$ are directly proportional to $\Delta\lambda/R$. As a consequence, the temperature variation in penetration depth of a superconductive sample can be determined by measuring changes in resonant frequency of a TDO circuit using the linear dependence in Eq.~\ref{eq:eq2}, where $G$ is a calibration constant depending on the dimensionality of the coil-sample setup which can be directly estimated by removing the sample from the coil at the lowest temperature~\cite{Prozorov00}.

\begin{equation}
\label{eq:eq2}
\Delta f = -\frac{G}{R} \Delta \lambda
\end{equation}

The sensitivity of the technique is strongly dependent on the filling factor of the sample, i.e. the ratio between the volume of the sample and that of the inductor. Typical TDO experiments use solenoid shape inductors however, for slab shaped specimens, the low filling factor can result in low resolution of the measurements. A more intuitive approach towards increasing the sensitivity is making use of planar inductors to probe plate like samples~\cite{Gevorgyan00,Gevorgyan04}. However, the complicated field  distribution of a single planar coil makes difficult to extract quantitative information. The direct proportionality between the frequency shift and penetration depth variation in Eq.~\ref{eq:eq2} was derived for slab like sample in uniform perpendicular field. In order to increase the filling factor, hence the sensitivity of our measurements, while providing a uniform perpendicular field in the region of the sample in normal state, we used a pair of planar inductors for our TDO setup.

\begin{figure}[t]
\includegraphics*[width=3.4 in, keepaspectratio=true]{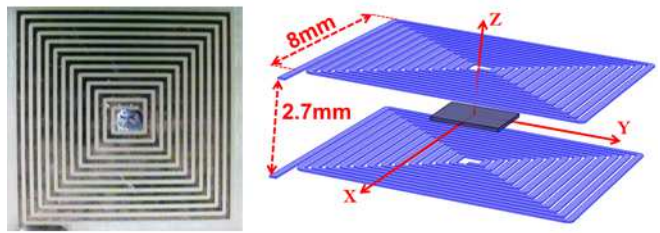}
\caption{\label{fig:fig1} (Color online) Left: Picture of one of the 8$\times$8 mm$^{2}$ flat coils with 3 turns/mm milled on a copper-clad laminate 1 oz. PCB board. Right: Spatial arrangement of the coils and sample. The setup is symmetric with respect to reflection across the z = 0 plane.}
\end{figure}

\begin{figure}[t]
\includegraphics*[width=3.4 in, keepaspectratio=true]{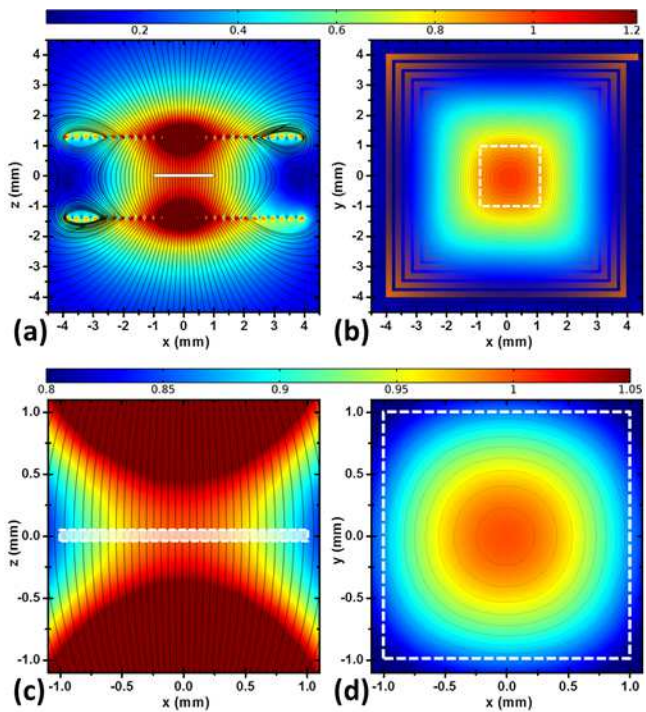}
\caption{\label{fig:fig2} (Color online) The simulated magnetic field distribution of our setup for the normal state of the sample. (a) Magnetic field lines and flux density distribution over the y$=$0 cross section of the setup (side view). (b) Flux density distribution over the z$=$0 cross section of the setup (top view). (c) Expanded view on the y$=$0 plane. (d) Expanded view on the z$=$0 plane. The white rectangles symbolize the domain of a 2$\times$2$\times$0.1 mm$^{3}$ sample. The color scale corresponds to the B field magnitude relative to its value in the center of the sample (0,0,0). }
\end{figure}

Pairs of planar rectangular spiral coils 8$\times$8 mm$^{2}$ in size, with 3 turns/mm, were milled on a copper-clad PCB board and connected in aiding parallel to form a sandwich configuration. The coils, separated by a 2.7 mm gap, are mirror-image of each other, and the sample is positioned midway with the $ab$ crystallographic plane parallel to the surface of the flat coils (Fig.~\ref{fig:fig1}). Considering the symmetry of our setup and the small thickness of the samples relative to the coil gap, the probing ac-field is parallel to the $c$-axis of the crystal ensuring that supercurrents are only induced in the $ab$ plane, thus the measured changes in resonant frequency are solely due to the variation in $\lambda_{ab}$.

To test for the uniformity of the field in the sample region, simulation were carried out for our specific coil-sample configuration using the COMSOL 4.2 Multiphysics software \cite{comsol}, a commercial finite element simulator. Figure~\ref{fig:fig2} depicts the simulated results obtained for the field lines and magnetic flux density distribution over the y$=$0 and z$=$0 cross sections of the setup in the normal state of the sample. The results confirm that the probing field from the coils is indeed perpendicular to the $ab$ surface of the sample (Fig.~\ref{fig:fig2}(c)) and that in a central rectangular region of dimensions comparable to the sample size, the magnitude of the field is homogeneous with $\sim 90\%$ uniformity (Fig.~\ref{fig:fig2}(d)).

Because of the strong dependence of the TDO's resonant frequency on temperature its inductor and electronic components were mounted on a special stage, thermally decoupled from the sample stage, and kept at a constant temperature of 3.7 K $\pm$ 0.001 K. The samples were mounted using Apiezon N grease on a 0.5 mm thick sapphire slab, attached to a copper block coupled to the mixing chamber. A ruthenium oxide thermometer in close proximity was used to measure the sample temperature. This way we were able to vary the sample temperature anywhere between 50 mK and 15 K, while the temperature of the oscillator remained constant, ensuring that the variations in the resonant frequency are exclusively caused by changes in the magnetic susceptibility of the sample. The resonant frequency of our empty oscillator is $f_0\sim$ 6 MHz, with a noise level lower than 0.5 Hz and with no detectable drift over the time period of a temperature run.
The relative variation in $\lambda_{ab}(T)$ was determined using Eq.~\ref{eq:eq2} where the effective dimension $R$ was calculated using the method described in Ref.\,\onlinecite{Prozorov00}.

Our TDO setup does not include a mechanism that would allow for physical extraction of the sample in-situ however, since the susceptibility of our samples in the normal state is negligible, the empty resonator frequency $f_0\approx f(T>T_C)$ thus $G$ can be calculated using $G\approx f(T=0)-f(T>T_C)$, where $f(T>T_C)$ is the frequency value when the sample temperature is above $T_C$. Considering values of $G$ as high as 200 kHz obtained for our specimens, from Eq.~\ref{eq:eq2}, we estimate the sensitivity of our setup for $\Delta\lambda_{ab}$ measurements to be around 1 nm.

\section{Results}

\begin{figure}[t!]
\includegraphics*[width=3.4 in, keepaspectratio=true]{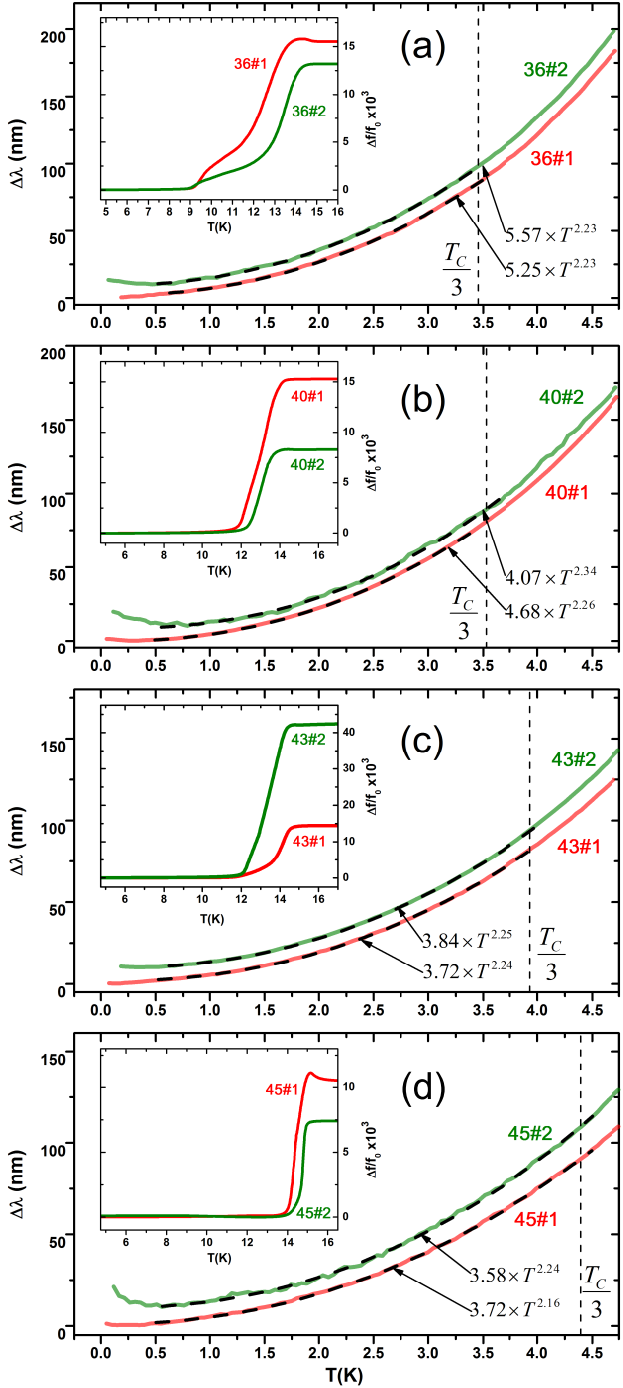}
\caption{\label{fig:fig3} (Color online) $\Delta\lambda_{ab}(T)$ (continuous lines) in Fe$_{1+y}$(Te$_{1-x}$Se$_{x})$  for the low temperature range in 2 different specimens for each nominal Se concentration namely (a)x=0.36, (b)x=0.40, (c)x=0.43, and (d)x=0.45. The dashed black lines are the representative allometric fits for each sample in the 0.5K$-T_c/3$ temperature range with the fitting parameters $A$ and $n$ shown. The curves have been offset by 10 nm for clarity. Inset: Relative frequency variations from TDO measurements for each sample. }
\end{figure}

The main panels of Fig.~\ref{fig:fig3} show the low temperature $\Delta\lambda_{ab}(T)$ for 8 samples discussed in this work, grouped by their nominal Se concentration with 2 samples for each. The insets show the relative TDO frequency change over the full measured temperature range\cite{note1}, including the transition at T$_{c}$. Broad transitions and additional humps can be observed in samples with 36\% Se concentration (see inset of Fig.~\ref{fig:fig3}(a)) which can be attributed to inhomogeneous superconducting transitions near the phase boundary where inhomogeneity is unavoidable ~\cite{Liu10,arxiv}. Nevertheless, the low temperature behavior of $\Delta\lambda_{ab}(T)$ is very similar to that of the other concentrations.

From the main panels of Fig.~\ref{fig:fig3}(a-d), we observe that when a temperature range between 0.5 K and about 0.3T$_{c}$ is used for analysis, like in most of the previous studies, $\Delta\lambda_{ab}(T)$ appears to have a well behaved power law dependence, $\Delta\lambda(T) = A T^{n}$, with the exponent $n$ ranging from 2.16 to 2.34 for all the samples, consistent with the previous reports in the same temperature range~\cite{Kim10,Klein10,Cho11}. As the penetration depth probes the density of excited low-energy quasi-particles, it is only at relatively low temperatures, the upper limit of 0.3T$_{c}$ being generally chosen, that conclusions about the gap symmetry can be inferred from its temperature dependence.

\begin{figure}[t!]
\includegraphics*[width=3.4 in, keepaspectratio=true]{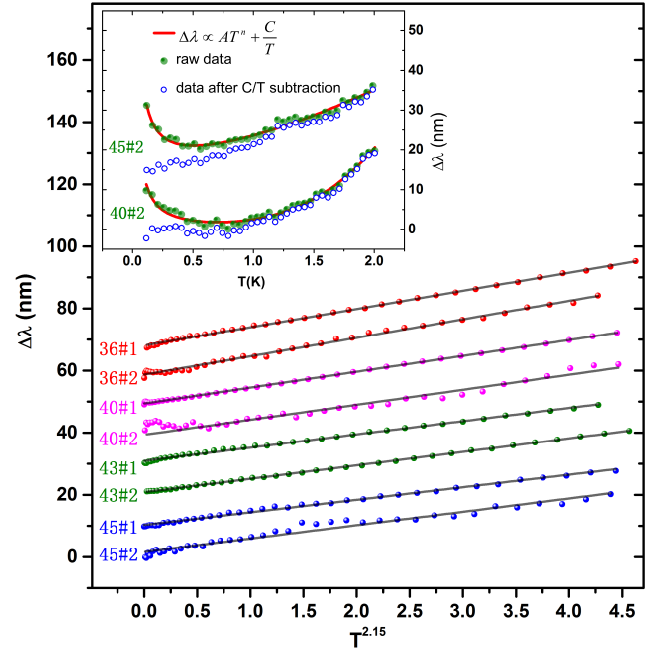}
\caption{\label{fig:fig4} (Color online) The relative variation of the in-plane penetration depth $\Delta\lambda_{ab}(T)$ data (points) at ultra-low temperatures for all 8 samples after subtracting the $C/T$ paramagnetic contribution as a function of $T^{2.15}$. The continuous lines are linear fits for the $T_{min} - $2K temperature range with the slope values of $A$ from Table~\ref{tab:table1}. The data for each sample has been shifted by 10 nm. Inset: the raw $\Delta\lambda_{ab}(T)$ data (filled spheres) and the data with the subtracted paramagnetic dependance (open circles)  for two samples, namely  40\#2 and 45\#2 (the data has been shifted by 20 nm). The continuous lines represent the $A T^{n}+ C/T$ fit of the raw data. }
\end{figure}

It can also be seen from Fig.~\ref{fig:fig3}(a-d) that in the limit of T$\rightarrow$0, most samples show an upturn of $\Delta\lambda(T)$. Similar upturn was also reported in a previous TDO work on Fe(Te$_{0.56}$Se$_{0.44})$ single crystals~\cite{Serafin10} and it was assigned to paramagnetic contribution from possible excess of Fe, occupying interstitial sites. In the insets of Fig.~\ref{fig:fig4} we show an example where penetration depth, from base temperature up to 2K, was fitted to a combination of power law and Curie contribution (Eq.~\ref{eq:eq4}).
\begin{equation}
\label{eq:eq4}
\Delta\lambda(T) = A T^{n} + \frac{C}{T},
\end{equation}
where $C$ is the Curie constant. The equation fits the data well for all samples and the resulting values for the free parameters $A$, $n$ and $C$ respectively, are summarized in Table~\ref{tab:table1}.

We would also like to mention that using a Curie-Weiss type equation for the magnetic contribution ($C/{(T-\Theta)}$), like in Ref.\,\onlinecite{Serafin10} did not improve significantly the quality of the fit. Following the same approach as in Ref.\,\onlinecite{Serafin10}, the parameter $C$ is given by
\begin{equation}
\label{eq:eq5}
C = -\frac{n_i \lambda_0 \mu_0 {\mu_e}^2}{6 k_B V_{cell}} \
\end{equation}
where $\mu_e$ is the effective magnetic moment of the paramagnetic ion. The resulting values of $C$, for the samples revealing an upturn at low temperature, span between 0.07 and 1.9 nm$\cdot$K, which would correspond to an average magnetic moment per unit cell value between 0.09$\mu_B$ and 0.5$\mu_B$ respectively (see Table~\ref{tab:table1}). We believe that the small excess iron $y$  could account for these low values of the magnetic moment and explain the paramagnetic behavior observed in most samples at low temperatures (see below for further discussions).

\begin{figure}[t!]
\includegraphics*[width=3.4 in, keepaspectratio=true]{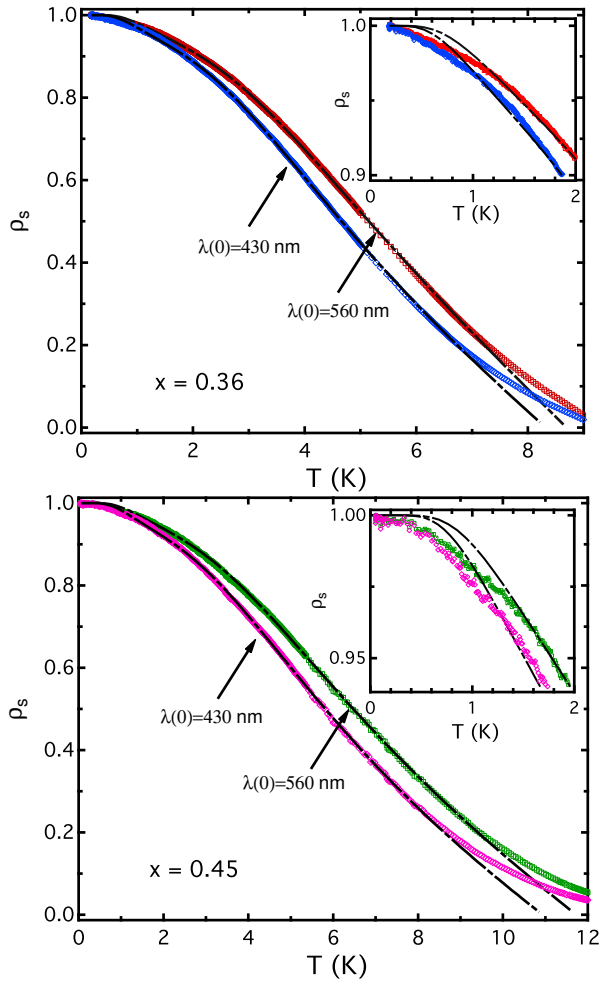}
\caption{\label{fig:fig5} (Color online) Superfluid density $\rho_{s}(T)$ in Fe$_{1.02}$Te$_{1-x}$Se$_{x}$ for the lowest Se doping x=36 (sample 36\#1, top) and highest Se doping x=45 (sample 45\#1, bottom) calculated from experimental data assuming two extreme values for $\lambda(0)$ reported in literature i.e. 430 nm\cite{Klein10} and 560 nm\cite{Kim10}. The dashed (black) lines illustrate the two-gap fit over the entire temperature range up to $T_C$. Inset: the low temperature region}
\end{figure}

In Table~\ref{tab:table1} we include the values of the parameters $A$ and $n$ from power-law fit $\Delta\lambda(T) = A\times T^{n}$ of the data below 2 K, after subtracting the magnetic contribution. Except for two samples (labeled 40\#2 and  45\#2), where the exponent was either significantly larger ($n\approx$~3.5), or lower ($n\approx$~1.5) than the rest, we found an average value of $n=2.15\pm0.25$. In the main panel of Fig.~\ref{fig:fig4} we plot $\Delta\lambda(T)$, after subtracting the magnetic contribution, as a function of $T^{2.15}$, where a well behaved linearity can be observed for the majority of our samples. We can therefore claim that the nearly quadratic temperature dependence of penetration depth in Fe$_{1.02}$(Te$_{1-x}$Se$_{x})$ is quite robust for all Se concentration. On one hand, the power-law behavior of $\Delta\lambda(T)$ is very similar to that observed in some of the iron pnictides~\cite{Gordon10}. On the other hand though, the fact that it persists clearly at all doping levels, including optimally doped, sets them apart from pnictides, where the low-energy excitations generally show behavior consistent with isotropic gap for optimal doping and with the existence of nodes for under/over doping~\cite{Reid10}. The values of the pre-factor $A$ for $n=2.15$ (Table~\ref{tab:table1}) also confirm the similarity between different Se concentrations. In each batch, the pre-factor has nearly the same value for most samples, $A=4.7\pm 1.2$ nm/K$^{2.15}$.  This result is also very different from pnictides, particularly the FeAs-122 family, where a much slower variation of penetration depth with temperature (i.e. lower value of $A$) was observed for optimally doped samples~\cite{Prozorov11}. One possible implication is that unlike in FeAs materials, the superconducting gap in Fe-chalcogenides may have the same structure for all Se concentrations, as we will discuss later.

Possible information about the superconducting gap(s) may be obtained by analyzing the superfluid density $\rho_{s}(T)=\left(\lambda(0)/\lambda(T)\right)^{2}$. In Fig.~\ref{fig:fig5} we show two examples, for $x$=0.36 and 0.45, corresponding to samples 36\#1 and 45\#1 respectively. The behavior of the superfluid density is strongly affected by the choice of $\lambda(0)$. Contrary to other Fe-based superconductors, previous reports of $\lambda(0)$ in Fe$_{1+y}$(Te$_{1-x}$Se$_{x})$ found very similar values for different values of $x$ and do not suggest a systematic evolution with Se concentration \cite{Bendele10,Biswas10,Kim10,Klein10}. We calculated $\rho_{s}(T)$ for two extreme values of $\lambda(0)$ reported in literature i.e. 430 nm and 560 nm from Ref.\,\onlinecite{Klein10} and Ref.\,\onlinecite{Kim10} respectively.

Similar to previous work~\cite{Bouquet01} on MgB$_{2}$, we consider the popular two-gap fit $\rho_{s} = \alpha\cdot\rho_{1}(\Delta_{1})+(1-\alpha)\cdot\rho_{2}(\Delta_{2})$, where $\rho_{1,2}$ are the superfluid density of the gap $\Delta_{1}$ and $\Delta_{2}$, respectively and $\alpha$ represents the relative contribution of the gaps~\cite{Bouquet01}. As it can be observed from Fig.~\ref{fig:fig5}, apparently the fit reproduces well the experimental data, and we obtain very similar behavior for all doping levels: $\Delta_{1}/\Delta_{2}\approx 3$ and $\alpha\approx 0.85$, i.e. the larger gap $\Delta_{1}$ contributes about 85\% to the superfluid density. We also found a systematic increase of $\Delta_{1}$ with Se concentration, by about 40\% at $x$=0.45 comparing with $x$=0.36, while $\Delta_{2}$ remained almost the same. These results are valid irrespective of the choice of  $\lambda(0)$ and while they may be qualitatively meaningful, there are serious issues with the fitting model. First, we mention that in all cases, both values of the gap resulted in lower than the BCS weak-coupling limit values of 1.76$k_{B}$T$_{c}$: $\Delta_{1}$ was about 1$k_{B}$T$_{c}$ and $\Delta_{2}\approx 0.3k_{B}T_{c}$. As it was previously discussed, for the iron pnictide superconductors this is clear indication that the model, which assumes that both gaps have BCS temperature dependence, with the same critical temperature, is not suitable for describing the superfluid density ~\cite{Kogan2009}. Second serious issue with this approach is that it fails to reproduce the experimental data at low temperature. We show two examples in the insets of Fig.~\ref{fig:fig5} and further mention that this was the case for the majority of samples.

We return now to the low temperature behavior of $\Delta\lambda_{ab}(T)$ and discuss possible implications on the structure of the superconducting gap(s). First, we recount that despite the effect of Se substitution on the critical temperature in Fe$_{1+y}$(Te$_{1-x}$Se$_{x})$, we did not find a significant evolution with Se content, neither in the exponent nor in the magnitude of $\Delta\lambda_{ab}(T)$. We propose that the nearly quadratic temperature dependence of penetration depth in Fe$_{1+y}$(Te$_{1-x}$Se$_{x})$ can be understood in terms of the pair-breaking by magnetic fluctuations at ($\pi$,0). Previous neutron scattering study~\cite{Thampy12} on samples from the same growth found that the ($\pi$,0) antiferromagnetic fluctuations, originating from interstitial Fe, persist even at the optimal doping level and freeze into cluster spin glass state at low temperature. Each spin cluster nucleates around interstitial Fe and involves more than 50 neighboring ions in the Fe plane. It was shown recently that such ($\pi$,0) magnetic correlations are sources of incoherent magnetic scattering, which gives rise to charge carrier localization in the normal state and to pair-breaking in the superconducting state~\cite{arxiv}. Given that all our samples have almost the same Fe excess of about 2\%, we believe that there are basically very similar sources of pair-breaking for all concentrations, which produces low energy excitations, hence power law dependence of penetration depth like discussed in Ref.\,\onlinecite{Mishra09}.

\begin{figure}[t!]
\includegraphics*[width=3.4 in, keepaspectratio=true]{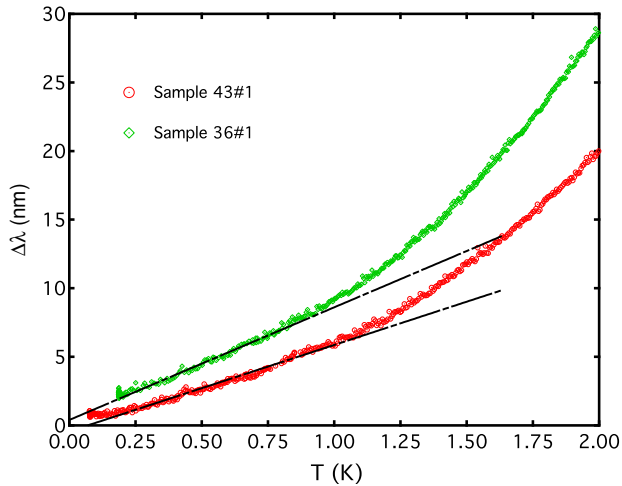}
\caption{\label{fig:fig6} (Color online) The relative variation of the in-plane penetration depth $\Delta\lambda_{ab}(T)$ raw experimental data (red points) for two samples with x=0.36 (36\#1) and x=0.43 (43\#1) for at low temperatures revealing a linear region.  }
\end{figure}

Additionally, we also suggest the possibility that at least one of the gaps is highly anisotropic, possibly nodal. It was shown theoretically~\cite{Vorontsov09} that for a superconducting gap with extended $s$-wave symmetry, without nodes, inter-band impurity scattering gives rise to a power-law temperature dependence of penetration depth $\Delta\lambda\propto T^{n}$, with an exponent as low as $n\approx$ 1.6. On the other hand, for an extended $s$-wave gap with nodes theory has shown~\cite{Mishra09} that ordinary disorder changes the otherwise linear behavior of $\Delta\lambda (T)$ into a power law with exponent $n\approx$ 2. The situation is similar to that of the cuprate superconductors, with d$_{x^{2}-y^{2}}$ gap-symmetry, where impurities give rise to a residual density of states\cite{Hirschfeld93}.

Therefore, both theoretical studies may be consistent with our quadratic temperature dependence of penetration depth observed experimentally. However, we emphasize that when the fit is restricted to very low temperatures, below 1K, $\Delta\lambda (T)$ is almost linear in some of the samples. This can be clearly observed from the superfluid density shown in the inset of Fig.~\ref{fig:fig5}, for 36\% Se concentration. In addition, we plot in Fig.~\ref{fig:fig6} the low temperature region of $\Delta\lambda (T)$ for this sample (36\#1) and for another one with 43\% Se (43\#1), i.e. closer to optimal doping. In both cases there is a clear linear region, albeit in a narrow temperature range. We also emphasize that these are two samples that did not show an upturn at low temperature (Table~\ref{tab:table1}), therefore ruling out possible artifacts due to the magnetic background subtraction. Given that for an $s_{\pm}$ gap symmetry without nodes, theoretical studies~\cite{Vorontsov09} have concluded that impurity scattering cannot generate a linear $\Delta\lambda (T)$, we believe that our data from Fig.~\ref{fig:fig6} is rather consistent with a nodal gap. For the other samples, impurities turn the otherwise linear penetration depth into a power-law, like discussed in Ref.\,\onlinecite{Mishra09}. Our finding appears to be consistent with the results from specific heat measurements under magnetic fields mentioned above~\cite{Zeng10} and with the theoretical model that predicts that gap on hole bands are fully gapped, while electron bands have nodal gaps or nodeless anisotropic gaps~\cite{Maier09,Kuroki09,Chubukov09,Wang09}.

\begin{table*}[t!]
\caption{\label{tab:table1} The values of the fitting parameters of Eq.~\ref{eq:eq2} for each sample and the corresponding magnetic moment.}
\begin{ruledtabular}
\begin{tabular}{lccccc}
Sample & $A[nm/K^{n}]$ & $n$ & $A(n=2.15)[nm/K^{2.15}]$ & $C[nm\cdot K]$ & $\mu_{unit cell}$ \\
\hline
\textit{36$\#$1} & 5.25 $\pm$0.07 & 2.23 $\pm$0.01 & 4.47 $\pm$0.03 & 0                &0\\
\textit{36$\#$2} & 6.11 $\pm$0.39 &  2.1 $\pm$0.08 & 5.93 $\pm$0.7  &  0.41 $\pm$0.06  &0.23$\mu_B$ \\
\hline
\textit{40$\#$1} & 4.99 $\pm$0.1  & 2.19 $\pm$0.03 & 5.11 $\pm$0.2  & 0.125 $\pm$0.012 &0.13$\mu_B$ \\
\textit{40$\#$2} & 1.81 $\pm$0.28 & 3.49 $\pm$0.21 & 4.83 $\pm$0.15 &  1.38 $\pm$0.11  &0.43$\mu_B$ \\
\hline
\textit{43$\#$1} & 3.72 $\pm$0.04 & 2.24 $\pm$0.01 & 4.25 $\pm$0.04 &  0               &0\\
\textit{43$\#$2} & 4.19 $\pm$0.11 &  2.2 $\pm$0.03 & 4.33 $\pm$0.02 &  0.31 $\pm$0.04  &0.2$\mu_B$ \\
\hline
\textit{45$\#$1} & 4.86 $\pm$0.18 & 1.88 $\pm$0.05 & 4.05 $\pm$0.05 &  0.07 $\pm$0.02  &0.096$\mu_B$ \\
\textit{45$\#$2} & 7.03 $\pm$1.05 & 1.47 $\pm$0.16 & 4.3 $\pm$0.14  &  1.93 $\pm$0.18  &0.5$\mu_B$ \\
\end{tabular}
\end{ruledtabular}
\end{table*}

\section{Conclusions}

To summarize, we have used a radio frequency tunnel diode oscillator technique to measure the in-plane London penetration depth in Fe$_{1+y}$(Te$_{1-x}$Se$_{x})$ single crystals with various Se concentrations down to temperatures as low as 0.05 K. We found that some samples show paramagnetic contribution below T $\approx$ 0.5 K. After subtracting the magnetic background, $\Delta\lambda(T)$ has a nearly quadratic temperature dependence for all Se concentrations. The magnitude of $\Delta\lambda(T)$ at low temperature is also very similar for all cases. Noticeably, we observed the presence of a region of linear $\Delta\lambda(T)$ in the limit of T$\rightarrow$0, both at low Se concentration and close to optimal doping. This is highly suggestive for the existence of nodes in the superconducting gap(s) of Fe$_{1+y}$(Te$_{1-x}$Se$_{x})$.
\newline \indent
\newline \indent

\section*{Acknowledgments}

The work was supported by the Louisiana Board of Regents through LA-SiGMA program under Award No. EPS-1003897 and the National Science Foundation under grant DMR-1205469.

\nocite{*}

\end{document}